\documentclass[aps, pre, twocolumn, superscriptaddress, nofootinbib, amsmath, amssymb]{revtex4-1}

\usepackage{graphicx}
\usepackage{dcolumn}
\usepackage{bm}
\usepackage{amsmath}
\usepackage[normalem]{ulem}

\newcommand{\rmd}{\ensuremath{\mathrm{d}}}
\newcommand{\tr}{\ensuremath{\mathrm{Tr}}}

\begin{document}

\title{The topological susceptibility of two-dimensional  $U(N)$  gauge theories}

\author{Claudio Bonati}
\email{claudio.bonati@df.unipi.it}

\author{Paolo Rossi}
\email{paolo.rossi@unipi.it}

\affiliation{Dipartimento di Fisica, Universit\`a di Pisa and INFN, Sezione di
Pisa, Largo Pontecorvo 3, 56127 Pisa, Italy}

\date{\today}

\begin{abstract}
In this paper we study the topological susceptibility of two-dimensional $U(N)$
gauge theories. We provide explicit expressions for the partition function and
the topological susceptibility at finite lattice spacing and finite volume. We
then examine the particularly simple case of the abelian $U(1)$ theory, the
continuum limit, the infinite volume limit, and we finally discuss the large
$N$ limit of our results.
\end{abstract}

\maketitle

\section{Introduction}

The study of $\theta$-dependence of QCD by means of lattice simulations has
been the subject of  several recent studies, mainly triggered by the possible
implications for axion physics \cite{Berkowitz:2015aua, Kitano:2015fla,
Borsanyi:2015cka, Trunin:2015yda, Bonati:2015vqz, Petreczky:2016vrs,
Frison:2016vuc, Borsanyi:2016ksw, Burger:2018fvb}. It is however well known
that Monte Carlo algorithms typically used in numerical simulations suffer from
a severe critical slowing down as the continuum limit is approached, with
autocorrelation times of topological observables that grow about exponentially
in the inverse of the lattice spacing \cite{Alles:1996vn, DelDebbio:2002xa,
DelDebbio:2004xh, Schaefer:2010hu}. This led to the development of new
algorithms, specifically devised to improve the sampling of topologically
nontrivial configuration \cite{Vicari:1992jy, Luscher:2011kk, Mages:2015scv,
Laio:2015era, Bietenholz:2015rsa, Frison:2016vuc, Borsanyi:2016ksw,
Hasenbusch:2017unr, Bonati:2017woi, Bonati:2017nhe, Bonati:2018blm,
Bonanno:2018xtd}. 

From a general point of view, it is very useful to have the possibility of
performing quantitative checks of the Monte Carlo results against exact ones.
This is obviously not possible in the general case, however simplified (toy)
models sometimes exist that are analytically soluble but still complicated
enough to be used as nontrivial test beds. In statistical physics the
two-dimensional Ising model is probably the most popular choice
\cite{FerrenbergLandauWong}, while in field theory two-dimensional lattice
gauge theories are the natural playground for tests of numerical simulations:
on one side they are computationally much cheaper than their four dimensional
counterparts, on the other side it is possible to determine many exact results
that may constitute  precise benchmarks for numerical results and
extrapolations.

The present paper is devoted to the extension of known analytic results
concerning two-dimensional $U(N)$ lattice gauge theories in the absence of a
$\theta$ term to the case when such a term is present, and more specifically to
the evaluation of the topological susceptibility, for  finite volumes $V$ and
for generic values of the coupling $\beta$. Finite volume results at fixed
coupling may be especially useful because they allow direct comparison with
simulations without the need of extrapolating to infinite volume and to the
continuum limit.  

The paper is organized as follows: Section \ref{sec:summary} is devoted to a
summary of known results, with special emphasis on finite lattices with
spherical and toroidal geometries. In Section \ref{sec:topology} we fix our
notation trying to make correspondence with previous literature as far as
possible, and we give our definitions for the density of topological charge and
for the topological susceptibility in $U(N)$ gauge theories, exploiting the
existence of the $U(1)$ subgroup. We present our general formulas for the
partition function in the presence of a $\theta$ term and for the topological
susceptibility, for generic values of $N$, $\beta$ and $V$, and for any genus
$g$ of the lattice manifold, showing explicitly that the periodicity of the
partition function for $2 \pi$ shifts of the $\theta$ parameter is preserved.
In Section \ref{sec:N1} we focus on the case $N=1$  where many closed-form
expressions can be explicitly found for generic values of $V$ and can be
compared with partial results already available in the literature.  Strong
evidence of precocious scaling by using a renormalized coupling is also
exhibited.  In Section \ref{sec:continuum} we analyze the (finite volume)
continuum limit of the model $\beta \rightarrow \infty$  in the presence of a
$\theta$ term.  In Section \ref{sec:infvolume} the infinite volume limit of the
topological susceptibility is discussed in detail.  Section \ref{sec:largeN} is
devoted to the study of the large $N$ limit in the infinite and finite volume
cases (with further evidence of precocious scaling) and to numerical checks of
our large $N$ results.

\section{A summary of known results}\label{sec:summary}

The finite volume lattice version of $U(N)$ gauge theories  most widely studied
in the literature is defined by the following  partition function \cite{Bars:1979xb}:
\begin{eqnarray}
& Z(N,\beta,P) =\int  e^{-S(N,\beta, P)} \prod_{l=1}^L  \rmd U_l \ , \\
& S(N,\beta, P) = -N\beta \sum_{p=1}^P  \tr (U_p+U_p^{\dag}), \label{eq:action}
\end{eqnarray}
where unitary $N \times N$ matrices $U_l$ are attached to the $L$ links of the
lattice and $U_p = \prod U_l$ are the ordered products of the link matrices
along any lattice plaquette. $\beta$ is the lattice 't Hooft coupling, whose
relationship with the standard (dimensionful) coupling\footnote{In the
following we will denote by $g$ also the genus of the manifold on which the theory
is defined; the meaning of $g$ should however be clear from the context.} is $N
\beta = 1/(g^2 a^2)$, where $a$ is the lattice spacing and the volume is given
by $V=P a^2$. The sum in Eq.~\eqref{eq:action} runs over all $P$ plaquettes,
while the integration $\rmd U_l$ involves all link variables and is performed
by using the Haar measure for the $U(N)$ group. Due to its crucial role we
recall that, when the integrand involves only functions of the eigenvalues
$\phi_i$ of the integration variable, the $U(N)$ Haar measure reduces to (see
e.g.  \cite{Drouffe:1983fv})
\begin{equation}
\rmd\mu (\phi) =   \Delta(\phi) {\bar \Delta } (\phi) \prod_{i=1}^N\frac{\rmd\phi_i}{2 \pi} \ ,
\end{equation}
where
\begin{equation}
\Delta(\phi) = \frac{1}{\sqrt{N!}}\epsilon_{j_1\cdot \cdot \cdot j_N}  e^{i \phi_1(N-j_1)}\cdot \cdot \cdot e^{i \phi_N(N-j_N)}\ .
\end{equation}

The peculiarity of two dimensional models consists in the possibility of
performing a change of integration variables (exploiting the invariance of the
Haar measure) in such a way that most nontrivial integrations involve directly
the plaquette matrices. That this is a feasible strategy can be understood, for
a two dimensional compact orientable manifold without boundary, by using the Euler
characteristic $2 -2g = S-L+P$, where $S$ is the number of sites (vertices) of
the lattice and $g$ is the genus of the lattice manifold.  The maximal number
of links that can be gauged away (maximal tree) is simply $S-1$ and therefore
the number of nontrivial integration variables $I$ is $I= L-S+1 = P-1 + 2g$. 

Two cases particularly useful for applications are the manifolds with the
topology of the sphere ($g=0$) and the manifolds with the topology of the torus
($g=1$). For the case $g=0$ we have $I = P-1$, implying that one of the
plaquette variables may be expressed as a function (actually the product) of
all other matrices; in this case one can easily prove the equivalence of these
models to the chiral chains of length $P$ (see also later in this section), in
order to use the results available for these systems \cite{Brower:1980rp,
Brower:1980vm}. For $g=1$ (the manifolds typically adopted in simulations) we
get $I = P+1$ and the independent variables may be chosen to be $P-1$
plaquettes and two other degrees of freedom (``torons'').  Integration over the
torons may be explicitly carried out \cite{Kiskis:2014lwa}, and the result
leads again to the possibility of expressing the last plaquette as the product
of all other variables. This procedure can be generalized without difficulties
also to the case of generic topology.

Without belaboring the details we only quote the final result, due to
Rusakov \cite{Rusakov:1990rs} (see also \cite{Kiskis:2014lwa} for the case of
the torus): the $\theta = 0$ partition function $Z^{(g)}(N,\beta, P)$
corresponding to a compact orientable lattice manifold of genus $g$ without
boundary is:
\begin{equation}\label{eq:Zg0}
Z^{(g)} (N, \beta, P)= \sum_r  d_r^{2-2g}  \Bigl[ \frac{\tilde{\beta}_r(N, \beta)}{d_r} \Bigr]^P,
\end{equation}
where $P > 1$,  the sum runs over all representations $r$ of $U(N)$, $d_r$ is the
dimension of the representation and \cite{Drouffe:1983fv} 
\begin{equation}
\tilde{\beta}_r (N, \beta)  = \int \chi_r (U) e^{N\beta (\tr U + \tr U^{\dag})}\rmd U\ ,
\end{equation}
with $\chi_r (U)$ the character of $r$. If the manifold is nonorientable the
partition function is always equal to 1, if fixed boundaries are present the
result depends on the holonomies associated to the boundaries
\cite{Rusakov:1990rs}. When the boundary holonomies are fixed to be trivial, one
obtains again Eq.~\eqref{eq:Zg0} and this is a possible way of proving the
equivalence of the spherical topology with chiral chains. We explicitly note
that, when writing expressions like Eq.~\eqref{eq:Zg0}, we must keep in mind
that the number of links belonging to each plaquette is not {\it a priori}
fixed, and for small values of $P$ it must  be large enough to ensure the
possibility of imposing boundary conditions compatible with the genus $g$ of
the lattice manifold. In particular for $P=2$ the plaquettes must be polygons
with at least $4\,g$ sides.

It is worth noticing that, due to the invariance properties of the measure,
a simple result may be obtained in the case $g=0$, $P=2$:
\begin{equation}
Z^{(0)} (N, \beta, 2) = {\tilde \beta}_0 (N, 2\beta) .
\end{equation}

We also recall that the continuum partition function in the case of a finite (dimensionless)
area $A =V/a^2$  can be obtained starting from the heat kernel action, corresponding  to
the replacement\footnote{There is sometimes confusion in the
literature on the numerical factor appearing in the exponent, which depends on
the conventions adopted in the action. We checked that Eq.~\eqref{eq:heat} is
the correct large $\beta$ limit of Eq.~\eqref{eq:detbessel}.} \cite{Drouffe:1983fv} 
\begin{equation}\label{eq:heat}
{\tilde \beta}_r (N, \beta)  \rightarrow  d_r  e^{-\frac{C_r}{4 N \beta}} \ ,
\end{equation}
where  $C_r$ is the quadratic Casimir in the $r$ representation and the result is 
\begin{equation}
Z^{(g)} (N, \beta, A) = \sum_r d_r^{2-2g}e^{-\frac{1}{4 N \beta} C_r A}\ .
\end{equation} 

The infinite volume limit of Eq.~\eqref{eq:Zg0} can be easily recovered in
different ways. For instance one may observe that when $P \rightarrow \infty$
it is consistent to choose an axial gauge condition, amounting to setting $U_l
= 1$ for all the links in the ``time'' direction of the lattice. Factorization
of the integrals in Eq.~\eqref{eq:action} follows trivially, implying a direct
relationship with the single plaquette model:
\begin{equation}\label{eq:singleplaqfact}
Z (N, \beta, P) \stackrel{P\rightarrow\infty}{\rightarrow} Z(N, \beta, 1)^P\ ,
\end{equation}
where
\begin{equation}
Z(N, \beta, 1) \equiv  \int \rmd U e^{N\beta (\tr U + \tr U^{\dag})} = {\tilde \beta}_0 (N, \beta)
\end{equation}
and the properties of the trivial representation ($d_0 = 1$ and $\chi_0(U) =
1$) have been exploited.  It is important to stress that the same result might
have been obtained by observing that the quantities $\tilde{\beta}_r (N,
\beta)$ can be explicitly computed for all values of $N$. Indeed by recalling
the definition of the modified Bessel functions of integer order
\begin{equation}\label{eq:bessel}
I_n (2 N \beta) = \frac{1}{2 \pi} \int_{-\pi}^{\pi}  e^{2  N\beta  \cos \phi \pm i n \phi} \rmd \phi
\end{equation}
it is possible to obtain the result \cite{Drouffe:1983fv}
\begin{equation}\label{eq:detbessel}
\tilde{\beta}_{\{l_j\}} (N, \beta) = \det\big(I_{l_j + i-j}(2 N\beta)\big)
\end{equation}
where the indices $l_1 \ge \cdot \cdot \cdot \ge l_N$
($l_i\in\mathbb{Z}$) parametrize the $U(N)$ representation; in
particular \cite{Bars:1979xb}
\begin{equation}\label{eq:detbessel0}
\tilde{\beta}_0 (N, \beta) = \det\big(I_{i-j}(2 N\beta)\big)\ .
\end{equation}
Noticing that $I_n(x) < I_0(x)$ for all $n \neq 0$ and for all finite real
values of $x$, it is easy to get convinced that 
\begin{equation}
\frac{\tilde{\beta}_r (N, \beta)}{d_r\,\tilde{\beta}_0 (N, \beta) } < 1 
\end{equation}
for all $r \neq 0$ and for all finite values of $\beta$. This observation
implies Eq.~\eqref{eq:singleplaqfact} and also that the convergence to the
infinite volume limit is exponentially fast for large values of $P$. 

As we mentioned in the introduction, many exact results have been obtained in
the past with regard to the large $N$ limit of many matrix models. For a
general review we refer to \cite{Rossi:1996hs}, quoting here only the existence
of a third order phase transition at $N=\infty$, first identified by Gross and
Witten \cite{Gross:1980he} and Wadia \cite{Wadia:1980cp}, the computation of
the first few $1/N$ corrections of the free energy \cite{Goldschmidt:1979hq},
the exact expression for the large $N$ eigenvalue distribution of the plaquette
variable for the single plaquette \cite{Gross:1980he} and for the chiral chains
with $P=3$ and $P=4$ \cite{Brower:1980rp, Brower:1980vm,Friedan:1980tu}, the
solution of the external field problem for all $N$ \cite{Brower:1980tb,
Brower:1980rp, Brower:1980vm, Brezin:1980rk} and the expectation value of $\det
U_p$ for the single plaquette model \cite{Rossi:1982vw}.

\section{Topological charge and susceptibility}\label{sec:topology}

The existence of a topological charge in two dimensional $U(N)$ gauge theories
is related to the existence of a $U(1)$ Abelian subgroup, that can be
parametrized  by a phase $\Phi$, related to the determinant of the $U(N)$
matrix by the relationship
\begin{equation}
\det U = e^{i  \Phi}\ .
\end{equation}
It is easy to get convinced that on a compact orientable lattice manifold
without boundaries the following property holds
\begin{equation}
\sum_{p=1}^P  \Phi_p = 0\, (\mathrm{mod}\, 2 \pi)\ ,
\end{equation}
where $\Phi_p$ is the  phase associated with the determinant of each plaquette variable.
Hence a simple definition for the topological charge density $q_p$ associated
with each plaquette is
\begin{equation}\label{eq:top_ch_dens}
q_p \equiv -\frac{i}{2 \pi} \ln \det U_p = -\frac{i}{2 \pi} \tr \ln U_p\ ;
\end{equation}
the total topological charge is $Q (N, P) = \sum  q_p$ and, because of the
above property of $\sum_p \Phi_p$, $Q$ can only take integer values.
Note that the second equality in Eq.~\eqref{eq:top_ch_dens} holds only
for an appropriate and $U_p$-dependent choice of the branch cuts. If however
the standard $[-\pi,\pi]$ branch is used (as will always be done in the
following), the two expressions for the topological charge are generically
different, but nevertheless the corresponding $\theta$-dependent partition
functions are the same.

By definition the (dimensionless) topological susceptibility $\chi_t (N, \beta, V)$ is
\begin{equation}
\chi_t (N, \beta, V) = \frac{a^2}{V} \Bigl[\langle Q^2\rangle -\langle Q\rangle^2 \Bigr],
\end{equation}
where the expectation values are to be computed at $\theta = 0$. The lattice
representation of $\chi_t$ follows trivially from the above results, recalling
that $V = a^2 P$, and simple parity arguments imply that $\langle Q\rangle = 0$,
therefore in practice we just have to compute $\langle Q^2\rangle/P$.  

The $\theta$-dependent partition function can be defined as
\begin{equation}
Z_{\theta} (N, \beta, P)  \equiv \int   e^{i \theta Q (N,P) } e^{-S(N,\beta, P)}
\prod_{l=1}^L  \rmd U_l  \ ,
\end{equation}
and in order to compute $Z_{\theta}(N, \beta, P)$ we can repeat and adapt
Rusakov's  procedure. Let us define  the quantities
\begin{equation}\label{eq:tildegamma}
\tilde{\gamma}_r (N, \beta, \theta)  \equiv  \int \chi_r (U) 
e^{\frac{\theta}{2 \pi }\tr \ln U + N\beta (\tr U + \tr U^{\dag})} \rmd U\ ,
\end{equation}
with the property that
\begin{equation}
\tilde{\gamma}_r (N, \beta, 0) = \tilde{\beta}_r (N, \beta).
\end{equation}
By choosing an appropriate gauge condition and performing the residual
nontrivial integrations we then obtain our general result for the $\theta$-dependent
partition function:
\begin{equation}
Z_{\theta}^{(g)} (N, \beta, P)= \sum_r  d_r^{2-2g}  \Bigl[ \frac{ \tilde{\gamma}_r(N, \beta, \theta)}{d_r} \Bigr]^P.
\end{equation}
Noticing that $\tr\ln U = i \sum_j \phi_j$, where $e^{i\phi_j}$ are the
eigenvalues of $U$, and defining the functions
\begin{equation}
\mathcal{I}_{\nu}(x) \equiv  \frac{1}{2\pi}\int_{- \pi}^{\pi} e^{i \nu \phi}  e^{x \cos \phi} \rmd \phi\ ,
\end{equation}
which for integer indices reduces to modified Bessel functions (see
Eq.~\eqref{eq:bessel}), we obtain the closed form expression
\begin{equation}
\tilde{\gamma}_{\{l_j\}} (N, \beta, \theta) =\det\left(\mathcal{I}_{l_j+i-j+\frac{\theta}{2 \pi }}(2 N\beta) \right)
\end{equation}
using arguments identical to those needed to prove Eqs.~\eqref{eq:detbessel}
and \eqref{eq:detbessel0} \cite{Bars:1979xb, Gross:1980he, Drouffe:1983fv}.

From this expression it follows that $\theta\to\theta+2\pi$ is equivalent to
$\{l_j\}\to\{l'_j\}$ where $l'_j=l_j+1$; as a consequence, when performing the
summation over all representations, each contribution appearing in $Z_{\theta+2
\pi}^{(g)} (N, \beta, P)$  has an identical counterpart  in the expression of
$Z_{\theta}^{(g)} (N, \beta, P)$, implying exact $2\pi$ periodicity in $\theta$
for all values of $g, N, \beta$ and $P$. We consider this to be a quite
nontrivial evidence for the correct normalization of the topological charge in
the two dimensional $U(N)$ gauge theories.

In order to simplify the notation it is convenient to introduce the weights
\begin{equation}\label{eq:weights}
w_r^{(g)}(N, \beta, P) =  d_r^{2-2g} \Bigl[ \frac{\tilde{\beta}_r(N, \beta)}{d_r} \Bigr]^P \Bigl[Z_0 (N, \beta, P)\Bigr]^{-1},
\end{equation}
with the property that $\sum_r w_r^{(g)} (N, \beta, P) = 1$.  Starting from the
formal expression for  the topological susceptibility:
\begin{equation}
\chi_t^{(g)}(N, \beta, P)  = -\frac{1}{P} \left.\frac{\partial^2 \ln Z_{\theta}^{(g)}(N, \beta, P)}{\partial \theta^2}\right|_{\theta =0}.
\end{equation}
it is then possible to represent $\chi_t^{(g)}(N, \beta, P)$ in the form
\begin{equation}\label{eq:chitopNV}
\begin{aligned}
&\chi_t ^{(g)}(N, \beta, P) = -\sum_r w_r^{(g)} (N, \beta, P)  \frac{\tilde{\gamma}''_r(N, \beta)}{d_r \tilde {\beta}_r (N, \beta) }  + \\
&-(P-1) \sum_r w_r^{(g)} (N, \beta, P) \Bigl[ \frac{\tilde{\gamma}'_r(N,\beta)}{d_r \tilde {\beta}_r (N, \beta) }\Bigr]^2
\end{aligned}
\end{equation}
where we have defined
\begin{equation}
\begin{aligned}
&\tilde{\gamma}'_r (N, \beta) \equiv \left.\frac{\partial \tilde{\gamma}_r (N, \beta, \theta)}{\partial \theta}\right|_{\theta=0}=\\
&=\int \chi_r (U) \Bigl( \frac{1}{2 \pi } \tr\ln U \Bigr) e^{N\beta (\tr U + \tr U^{\dag})} \rmd U\ ,
\end{aligned}
\end{equation}
and
\begin{equation}
\begin{aligned}
&\tilde{\gamma}''_r(N, \beta) \equiv \left.\frac{\partial^2{\tilde \gamma}_r (N, \beta, \theta)}{\partial \theta^2}\right|_{\theta=0}= \\
&= \int \chi_r (U) \Bigl(\frac{1}{2 \pi } \tr \ln U \Bigr)^2 e^{N\beta (\tr U + \tr U^{\dag})}\rmd U\ ,
\end{aligned}
\end{equation}
which can be rewritten as sums of determinants involving modified Bessel
functions and related functions (see Sec.~\ref{sec:infvolume} for more details on the
simplest case). In the derivation of  Eq.~\eqref{eq:chitopNV} we have also
exploited the fact that $\langle Q\rangle$ vanishes at $\theta=0$, which is
equivalent to
\begin{equation}
\sum_r w_r^{(g)} (N, \beta, P)  \frac{\tilde{\gamma}'_r(N,\beta)}{d_r \tilde {\beta}_r (N, \beta) } = 0\ .
\end{equation}
The proof of this identity rests on the cancellation of the contributions
coming from each representation $r$ (associated to $\{l_j\}$) and its conjugate
representation $r^*$ (associated to $\{-l_{N+1-j}\}$), indeed 
\begin{equation}
\begin{gathered}
d_{r^*} = d_r\ ; \quad  \tilde{\beta}_{r^*}(N, \beta) = \tilde{\beta}_r (N, \beta)\ , \\
\tilde{\gamma}_{r^*} (N, \beta, \theta) = \tilde{\gamma}_r (N, \beta, -\theta) \ ,
\end{gathered}
\end{equation}
from which it follows $\tilde\gamma'_{r^*} (N, \beta) = - \tilde \gamma'_r (N, \beta)$.

\begin{figure}[t] 
\centering 
\includegraphics[width=0.9\columnwidth, clip]{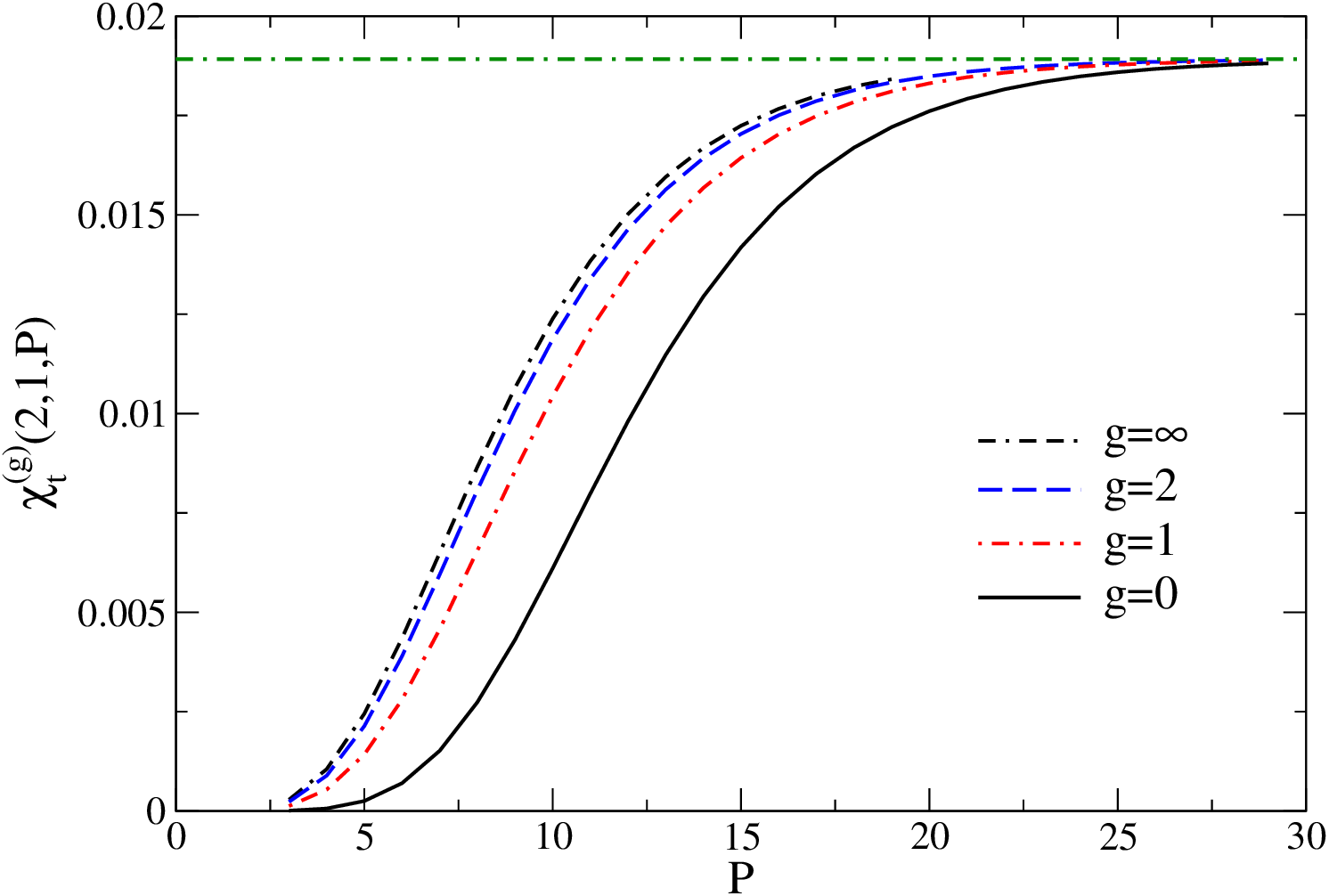}
\caption{Behaviour of the $U(2)$ topological susceptibility as a function of
the dimensionless volume $P$ for the value $\beta=1$ of the coupling. Results
are shown for three different topologies, corresponding to $g=0, 1$ and $2$;
the horizontal line denotes the asymptotic $P\to\infty$ value computed by using
Eq.~\eqref{eq:chiNinfiniteV}.} 
\label{fig:u2_finiteV}
\end{figure}

By the same arguments applied in the previous Section, and observing that
$\tilde{\gamma}'_0 (N, \beta) = 0$ for obvious symmetry reasons, we may
conclude that also the convergence of the topological susceptibility to its
infinite volume limit is exponentially fast. An example of the finite volume
behaviour of the topological susceptibility is shown in
Fig.~\ref{fig:u2_finiteV} for the $U(2)$ case.

A peculiar property of the case $g=0$, $P = 2$ (the two-link chiral chain) is
\begin{equation}
\chi_t^{(0)}(N, \beta, 2) = 0,
\end{equation}
implied by the trivial relationship $\tr \ln U + \tr \ln U^{\dag} = 0$.

It is also quite  interesting to study the limit $g\to\infty$ of the theory. Since
$g$ appears in the exponent of $1/d_r$ in the weights Eq.~\eqref{eq:weights},
all representations with $d_r>1$ disappear as $g\to\infty$ and only the
representations labelled by $l_1=l_2\cdots=l_N$ give a finite contribution in this limit.
The relevant representations are therefore identified by a single index $l$,
running from $-\infty$ to $+\infty$, and the partition function is simply
\begin{equation}\label{eq:largeg}
Z_{\theta}^{(\infty)}(N,\beta,P) = \sum_l \left[\tilde{\gamma}_l (N, \beta, \theta)\right]^P
\end{equation}
where the explicit form of the $U(N)$ characters has been used (see \cite{Drouffe:1983fv}) and
\begin{equation}\label{eq:largeg2}
\tilde{\gamma}_l (N, \beta, \theta) = \tilde{\beta}_0(N,\beta)\left\langle (\det U)^{l+{\theta \over 2 \pi }}\right\rangle\ ,
\end{equation}
where the average $\langle \quad\rangle$ stands for the average in the single
plaquette model at $\theta=0$. The topological susceptibility of the $g=\infty$
theory is shown in Fig.~\ref{fig:u2_finiteV} for the $U(2)$ case. Further
aspects of the large $g$ behaviour will be discussed in
Sec.~\ref{sec:continuum} and Sec.~\ref{sec:largeN}.

To better understand the form of Eq.~\eqref{eq:chitopNV} it is convenient to
further generalize the problem, by introducing plaquette dependent lattice
coupling and $\theta$ angle. It is immediate to verify that the Rusakov result can
be generalized to this case and the partition function becomes
\begin{equation}
Z_{\vec{\theta}}^{(g)} (N, \vec{\beta}, P)= \sum_r  d_r^{2-2g}  \prod_{p=1}^{P} \frac{\tilde{\gamma}_r(N, \beta_p, \theta_p)}{d_r}\ .
\end{equation}

We can now write a formal expression for the two-point correlation function of the topological charge by using
\begin{equation}
\langle q_i q_j\rangle^{(g)}(N,\beta,P)=-\left.\frac{\partial^2}{\partial\theta_i\partial\theta_j}
\ln Z_{\vec{\theta}}^{(g)} (N, \vec{\beta}, P)\right|_{\substack{\theta_p=0 \\ \beta_p=\beta}}
\end{equation}
and it is simple to verify that $\langle q_iq_j\rangle^{(g)}(N,\beta,P)$ has the form
\begin{equation}
\begin{aligned}
\langle q_iq_j\rangle^{(g)}(N,\beta,P)&=c_1^{(g)}(N,\beta,P)\delta_{ij}+\\&+c_2^{(g)}(N,\beta,P)(1-\delta_{ij})\ ,
\end{aligned}
\end{equation}
which expresses the fact that in two dimensions the correlator $\langle q_i
q_j\rangle^{(g)}$ takes just two values.  These values are obviously related to
the expressions appearing in Eq.~\eqref{eq:chitopNV}, that can indeed be
rewritten in the form
\begin{equation}
\chi_t^{(g)}(N,\beta,P)=c_1^{(g)}(N,\beta,P)+(P-1)c_2^{(g)}(N,\beta,P)\ .
\end{equation}
This is nothing but the general relation between the susceptibility and
the two point function, written in the case in which $\langle q_iq_j\rangle^{(g)}$
assumes only two values. Since $\tilde{\gamma}'_0 (N, \beta) = 0$, it is simple
to show that $c_2$ goes to zero exponentially in $P$ (the dimensionless volume)
as the thermodynamic limit is approached; in this limit the two point function
of the topological charge reduces to a $\delta$ function.

\section{The case $N=1$}\label{sec:N1}

In the purely Abelian case $N=1$ many simplifications occur, due to the
commutativity of the matrices. In particular there is no dependence on the
genus of the manifold, as one can easily see by noticing that all the
representations have dimension 1.

\begin{figure}[t] 
\centering 
\includegraphics[width=0.9\columnwidth, clip]{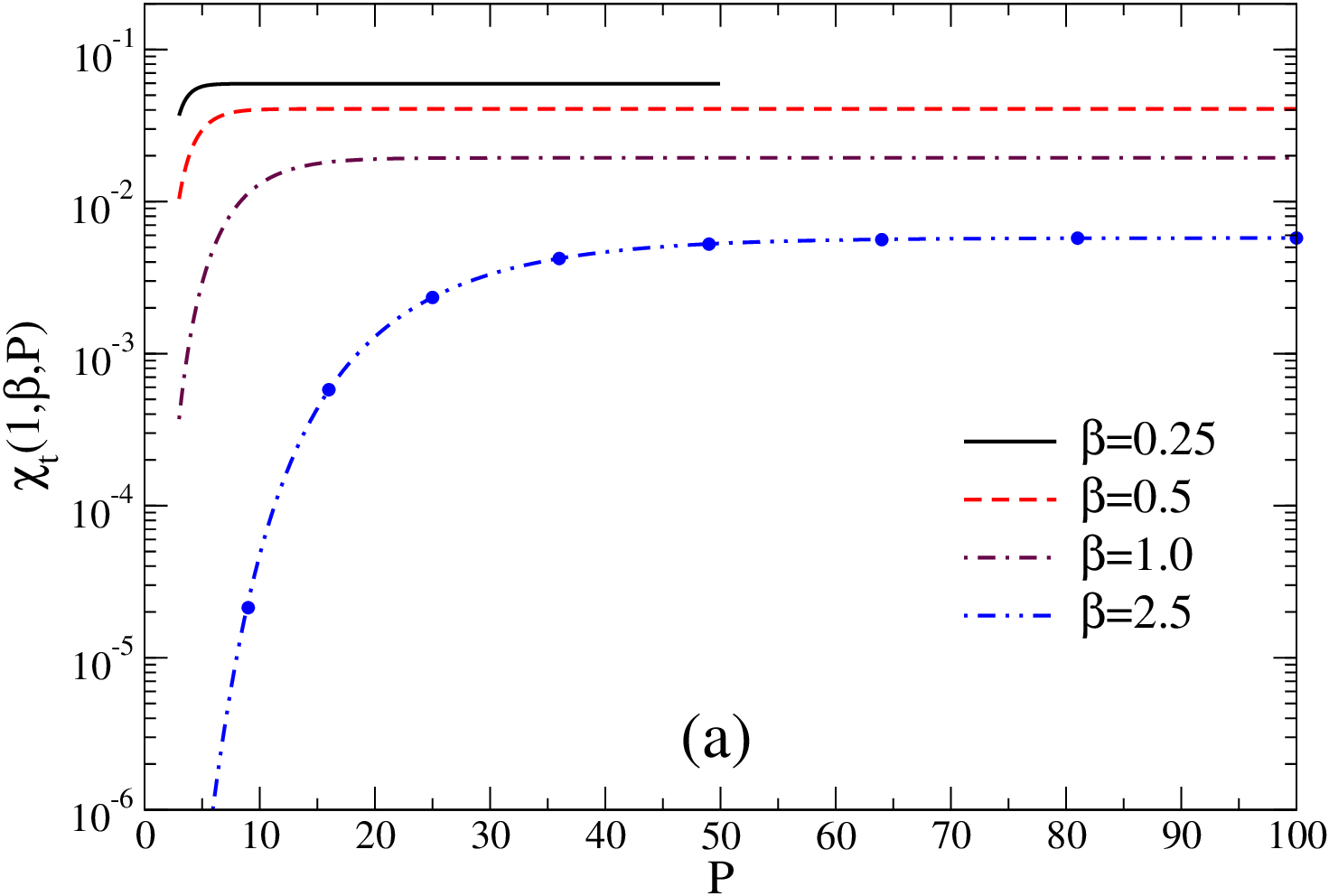}
\includegraphics[width=0.9\columnwidth, clip]{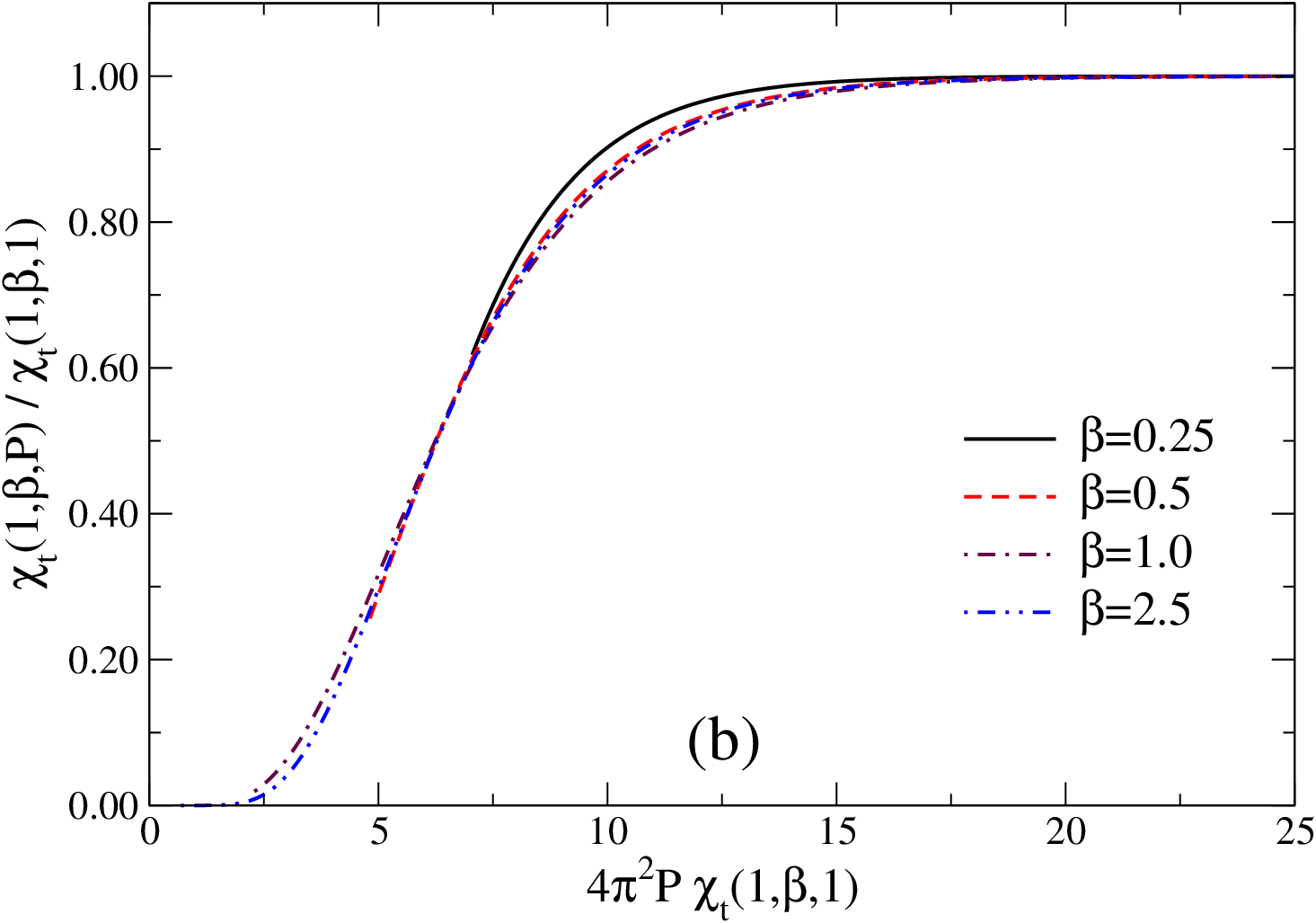}
\caption{(a) Behaviour of the $U(1)$ topological susceptibility as a function
of the coupling $\beta$ and of the dimensionless volume $P$. For comparison
points obtained by using numerical lattice simulations are also shown in the
case $\beta=2.5$ (error-bars are smaller than symbols). (b) Same data as in the
upper panel but with quantities normalized by using
$\chi_t(1,\beta,1)$. 
}
\label{fig:u1_finiteV}
\end{figure}

The topological charge density is simply $q_p = \frac{\phi_p}{2 \pi}$ (where
$\phi_p$ is the Abelian phase of the plaquette) and the character of the $n$-th
representation of $U_p$ is just $e^{i n \phi_p}$. As a consequence one may
compute directly the $\theta$ dependent  partition function on a finite lattice
obtaining
\begin{equation}\label{eq:Zu1}
Z_{\theta} (1, \beta, P) = \sum_{n = - \infty}^{+\infty} \Bigl[  \mathcal{I}_{n+\frac{\theta}{2 \pi}}  \Bigr]^P.
\end{equation}
The $U(1)$ weights are simply
\begin{equation}
w_n (1, \beta, P) = {\bigl[I_n (2 \beta)\bigr]^P \over \sum_n \bigl[I_n (2 \beta)\bigr]^P}
\end{equation}
The resulting expression for the finite volume topological susceptibility is then
\begin{equation}
\begin{aligned}
&\chi_t (1, \beta, P) =  -\sum_n  w_n(1, \beta, P)  {\mathcal{I}''_n(2 \beta) \over I_n (2 \beta) } +\\
&- (P-1) \sum_n w_n (1, \beta,P) \Bigl[ {{\mathcal{I}'_n (2 \beta) \over I_n(2 \beta)} \Bigr]^2 }\ ,
\end{aligned}
\end{equation}
where we introduced the auxiliary functions 
\begin{equation}
\begin{aligned}
\mathcal{I}'_n (x) & \equiv   \left. \frac{1}{2 \pi}\frac{\partial}{\partial \nu} \mathcal{I}_{\nu}(x)\right|_{\nu=n} =\\
&= \frac{i}{2 \pi} \int_{- \pi}^{\pi}  \frac{\phi}{2 \pi } e^{i n \phi + x \cos \phi} \rmd \phi
\end{aligned}
\end{equation}
\begin{equation}
\begin{aligned}
\mathcal{I}''_n (x) &\equiv  \frac{1}{(2 \pi)^2} \left.\frac{\partial^2}{\partial \nu^2} \mathcal{I}_{\nu}(x)\right|_{\nu=n} =\\
&=- \frac{1}{2 \pi} \int_{- \pi}^{\pi}  \left(\frac{\phi}{2 \pi}\right)^2 e^{i n \phi + x \cos \phi}\rmd \phi \ .
\end{aligned}
\end{equation}
The typical behaviour of $\chi_t(1,\beta,P)$ as a function of $\beta$ and $P$
is shown in Fig.~\ref{fig:u1_finiteV}(a). 
In Fig.~\ref{fig:u1_finiteV}(b) one may observe the precocious
scaling exhibited by the ratio $\chi_t(1,\beta, P)/\chi_t(1,\beta,1)$, when we parametrize the dependence on the
coupling by means of the combination $4\pi^2P\chi_t(1,\beta,1)$, corresponding
to a physical dimensionless quantity in the continuum limit (where it takes the
asymptotic value $\frac{P}{2\beta}$). Precocious scaling by use of renormalized
couplings was observed in a different context in references \cite{tadpole,
Lepage:1992xa}.

The finite volume continuum limit of the $\theta$ dependent  partition function in the $U(1)$ case is
\begin{equation}
Z_{\theta} \big(1, \frac{A}{2\beta}\big) = \sum_n e^{-\frac{A}{4 \beta} \left(n+\frac{\theta}{2 \pi} \right)^2}\ ,
\end{equation}
where we dropped the $\theta-$independent multiplicative factor $I_0(2\beta)^P$.
A corresponding expression for the topological susceptibility can easily be obtained, which can be written in the form 
\begin{equation}
\frac{\chi_t (1, \beta, A)}{\chi_t(1,\beta,1)}=1+2X\frac{\partial}{\partial X} \ln Z_0(1,X)\ ,
\end{equation}
where $X=\frac{A}{2\beta}$. In order to compare with continuum results (see e.g.  \cite{Cao:2013na}), it
must be kept in mind that $g^2=2e^2$ in the $U(1)$ case to preserve the
canonical normalization of the fields (see also the note at the end of Section \ref{sec:summary}).

In the infinite volume limit the dominant term of the sum in Eq.~\eqref{eq:Zu1} is
the one corresponding to the minimum value of $n+\frac{\theta}{2\pi}$, and we thus
see the emergence of a multi-branched structure
\begin{equation}
Z_{\theta} (1, \beta, P) \stackrel{P\rightarrow\infty}{\rightarrow} 
\left[\mathcal{I}_{\frac{\theta\,\mathrm{mod}\, 2\pi}{2 \pi}}(2 \beta)\right]^P\ , 
\end{equation}
with the partition function being non-analytic at the odd multiples of $\pi$.
This phenomenon persists also when considering the infinite volume limit of the 
continuum version of the model discussed above.
The presence of these first order transition points prevents a simple
factorization of the form Eq.~\eqref{eq:singleplaqfact} from being applicable for
generic $\theta$ values, indeed a \emph{naive} application of factorization would give
\begin{equation}
Z_{\theta}(1, \beta, P) \stackrel{P\rightarrow\infty}{\rightarrow} 
\left[\mathcal{I}_{\frac{\theta}{2 \pi}}(2 \beta)\right]^P\ , 
\end{equation}
which is non periodic in $\theta$. It is however important to stress
that, as far as we consider $-\pi\le\theta\le \pi$, all the expressions
obtained by using the single plaquette model correctly describe the
$P\to\infty$ limit of the $P$-plaquette model.
In particular the infinite volume topological susceptibility is given by
\begin{equation}
\chi_t (1, \beta, \infty) =-\frac{\mathcal{I}''_0(2 \beta)}{I_0 (2 \beta)}.
\end{equation}

\section{The continuum limit}\label{sec:continuum}

The continuum limit of two dimensional $U(N)$  gauge theories is simply the
limit $\beta \rightarrow \infty$ because the coupling $g^2$ is dimensionful and
therefore the above limit is the same as the limit $a^2 \rightarrow 0$.  By
generalizing the arguments  that lead to  Eq.~\eqref{eq:detbessel} we may
obtain the following representation for the functions  $\tilde{\gamma}_r(N,
\beta, \theta)$ appearing in $Z_{\theta} (N, \beta, P)$:
\begin{equation}
\begin{aligned}
 \tilde{\gamma}_{\{l_j\}}(N, \beta, \theta) &= \int \det \left[ e^{i \phi_j(l_j+i-j)} \right] \times \\
& \times e^{i {\theta \over 2 \pi}\sum_j \phi_j} e^{2N\beta \sum_j \cos \phi_j} \prod_j {\rmd \phi_j \over 2 \pi }\ .
\end{aligned}
\end{equation}

In the $\beta \rightarrow \infty$ limit one may replace $\cos \phi_j$ with $1 -
{1 \over 2}\phi_j^2$ and perform the resulting gaussian integration, thus
obtaining
\begin{equation}
\tilde{\gamma}_{\{l_j\}}(N, \beta, \theta) \rightarrow A (N, \beta) \det 
\left[ e^{-\frac{1}{4N\beta}\left(l_j +i -j + \frac{\theta}{2 \pi}\right)^2} \right],
\end{equation}
where the common factor $A(N, \beta)$ does not depend on $\theta$.

A few straightforward manipulations allow to represent the above result in the form
\begin{equation}
\begin{aligned}
\tilde{\gamma}_{\{l_j\}}(N, \beta, \theta) \rightarrow & A(N, \beta)  \det \left[ e^{-\frac{1}{4N\beta}(l_j +i -j)^2} \right] \times \\
& \times e^{{\theta \over \pi} \sum_j  l_j+ N\left(\frac{\theta}{2 \pi}\right)^2} .
\end{aligned}
\end{equation}
The determinant can be computed in the limit $\beta \rightarrow \infty$, obtaining the result
\begin{equation}
\det \left[ e^{-{1 \over 4N\beta}(l_j +i -j)^2} \right] \rightarrow B(N, \beta)\, d_{\{l_j\}}\, e^{-{1 \over 4N\beta}C_{\{l_j\}}},
\end{equation}
where $B(N, \beta)$ is another common factor independent of $\theta$, and
it is possible to verify that the product $A(N, \beta) B(N, \beta)$ is nothing but the asymptotic form of 
$\tilde{\beta}_0 (N, \beta)$ in the large $\beta$ limit, hence it  is a lattice artifact that can be ignored
 when analyzing the continuum properties of the model.

We recall that $C_{\{l_j\}}$ is the quadratic Casimir of the representation, as
expected from the $\theta =0$ result Eq.~\eqref{eq:heat}.  We report here, for
the convenience of the reader, the known explicit form of $C_{\{l_j\}}$ and
$d_{\{l_j\}}$:
\begin{equation}
\begin{aligned}
& C_{\{l_j\}}=\sum_{i=1}^N l_i(l_i-2i+N+1) \\
& d_{\{l_j\}}=\prod_{i>j}\left(1-\frac{l_i-l_j}{i-j}\right)\ .
\end{aligned}
\end{equation}

The continuum limit of the partition function on a manifold with
(dimensionless) area $A/(N\beta)=g^2V$ is therefore
\begin{equation}\label{eq:continuumpart}
Z_{\theta}^{(g)} \big(N, \frac{A}{2\beta}\big) = \sum_{\{l_j\}} d_{\{l_j\}}^{2-2g} e^{-{A \over 4N\beta}\big[C_{\{l_j\}}+ {\theta \over \pi }\sum_j l_j + {N \over 4 \pi^2}\theta^2 \big] }.
\end{equation}
and the continuum limit of the weights defined in Eq.~\eqref{eq:weights} is
\begin{equation}\label{eq:cont_weights}
w_{\{l_j\}}^{(g)} \big(N, \frac{A}{2\beta}\big) = d_{\{l_j\}}^{2-2g} e^{-{A \over 4N\beta} C_{\{l_j\}} } \Bigl[Z_0^{(g)}  \big(N, \frac{A}{2\beta}\big) \Bigr]^{-1}.
\end{equation}
An immediate consequence of the above results is the possibility of evaluating the
finite volume continuum limit of the topological susceptibility: 
\begin{equation}
\chi_t^{(g)}(N, \beta, A) = {1 \over 8 \pi^2 \beta}\Big[1 - {A \over 2\beta}
\sum_{\{l_j\}} w_{\{l_j\}}^{(g)} \big(\sum_j \frac{l_j}{N} \big)^2\Big]
\end{equation}
which in the infinite volume limit does not depend on the genus and becomes simply 
\begin{equation}
\chi_t^{(g)}(N, \beta, \infty) = {1 \over 8 \pi^2 \beta}\ ,
\end{equation}
for all $N$, because $w_r^{(g)} (N, \beta, A) \rightarrow \delta_{r,0}$ when $A \rightarrow \infty$.

It is important to note that the continuum expression for the partition
function is consistent with the previously proven periodicity in
$\theta$ with period $2 \pi$ of the partition function. Let's focus on the exponents appearing in
Eq.~\eqref{eq:continuumpart} and notice that
they can be rewritten in the form
\begin{equation}
\begin{aligned}
& C_{\{l_j\}}+{\theta \over \pi }\sum_j l_j + {N \over 4 \pi^2}\theta^2 = \\
& = \sum_j \left[\left(l_j+ {\theta \over 2 \pi}\right)^2 + (N+1 -2 j)\left(l_j+ {\theta \over 2 \pi}\right) \right]\ ;
\end{aligned}
\end{equation}
also in the continuum $\theta\to\theta+2\pi$ is thus equivalent to
$\{l_j\}\to\{l'_j\}$ where $l'_j=l_j+1$. Since $d_{\{l'_j\}} = d_{\{l_j\}}$ the
periodicity of the continuum partition function Eq.~\eqref{eq:continuumpart}
follows as in Sec.~\ref{sec:topology}.

The continuum version of the $g \rightarrow \infty$ limit is simply
\begin{equation}\label{eq:Zginf}
Z_{\theta}^{(\infty)} \big(N, \frac{A}{2\beta}\big) = \sum_l e^{-{A \over 4\beta}(l+{\theta \over 2\pi})^2},
\end{equation}
and one may appreciate that it turns out to be independent of $N$ and therefore
coincident with the continuum  version of the $U(1)$ model. However we notice
that, contrary to naive expectations, the finite volume continuum limit will
not in general coincide with its $U(1)$ value, and will depend on $N$ and $g$,
with the notable exception of the large $N$ limit, to be discussed in
Sec.~\ref{sec:largeN}.

The properties of the finite volume continuum limit will be discussed in detail in a forthcoming publication.

\section{The infinite volume limit}\label{sec:infvolume}

\begin{figure}[t] 
\centering 
\includegraphics[width=0.9\columnwidth, clip]{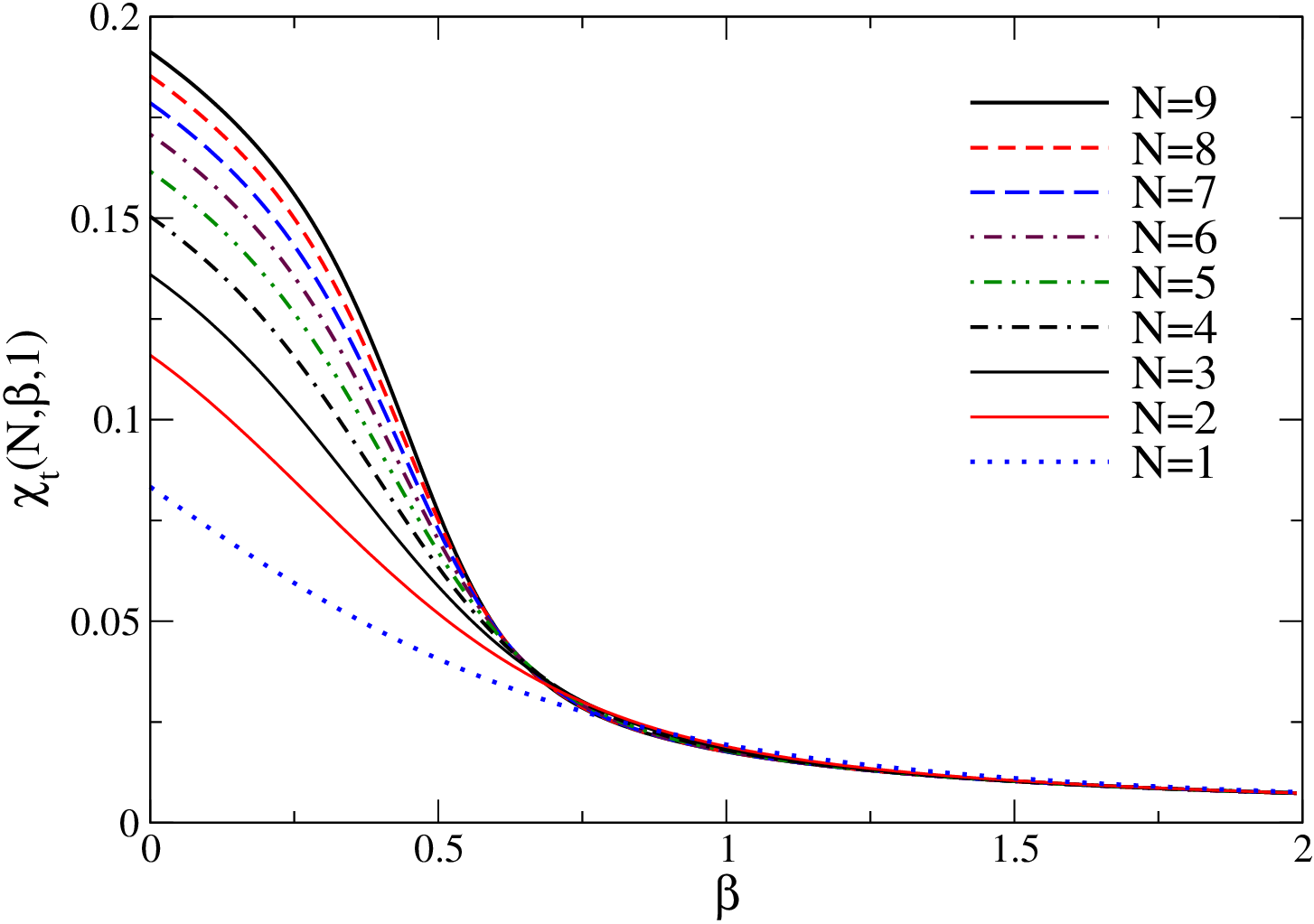}
\caption{Behaviour of the infinite volume topological susceptibility for $U(N)$
with $N<10$ and $0\le \beta\le 2$, computed using Eq.~\eqref{eq:chiNinfiniteV}.}
\label{fig:chi_vinfty}
\end{figure}

We assume in this section $-\pi\le \theta\le \pi$ (see the discussion in
Sec.~\ref{sec:N1}), in order to exploit the large volume factorization also at
$\theta\neq 0$, obtaining for all genuses
\begin{equation}
Z_{\theta}^{(g)} (N, \beta, P) \stackrel{P\rightarrow\infty}{\rightarrow} Z_{\theta}(N, \beta, 1)^P\ ,
\end{equation}
where
\begin{equation}
\begin{aligned}
Z_{\theta}(N, \beta, 1) & \equiv  \int e^{\frac{\theta}{2 \pi }\tr \ln U + N\beta(\tr U + \tr U^{\dag})}\rmd U = \\
& =\tilde{\gamma}_0 (N, \beta, \theta) \ .
\end{aligned}
\end{equation}
Computing the infinite volume topological susceptibility thus amounts to
evaluating the quantity
\begin{equation}\label{eq:chiNinfiniteV}
\chi_t (N, \beta, 1) = -\frac{\tilde{\gamma}''_0 (N, \beta)}{\tilde{\beta}_0(N, \beta)}\ ,
\end{equation}
where we exploited the fact that  $\tilde{\gamma}'_0 (N, \beta) = 0$ and the property
\begin{equation}
w_r^{(g)} (N, \beta, P) \rightarrow \delta_{r,0}
\end{equation}
in the limit $P \rightarrow \infty$.
$\tilde{\gamma}''_0$ may be evaluated starting from 
\begin{equation}
\tilde{\gamma}''_0 = -\int d\mu (\phi) \left(\sum_i\frac{\phi_i}{2 \pi}\right)^2 e^{2 N \beta \sum_i  \cos \phi_i}\ ,
\end{equation}
and it can be seen (using again arguments analogous to those of
\cite{Bars:1979xb,Gross:1980he, Drouffe:1983fv}) that $\tilde{\gamma}''_0$ may
be expressed as the sum of the $N^2$ determinants obtained from $\det I_{i-j}(2
N \beta)$ by replacing one of the lines with $\mathcal{I}''_{i-j}(2 N \beta)$
and two different lines with $\mathcal{I}'_{i-j}(2 N \beta)$. Using these
expressions it is straightforward to numerically compute $\chi_t(N,\beta,1)$
and in Fig.~\ref{fig:chi_vinfty} we show the results obtained for $N<10$ and
$0\le \beta\le 2$; two different regimes are clearly visible in this figure,
which will be discussed in depth in Sec.~\ref{sec:largeN}.

\section{The large $N$ limit}\label{sec:largeN}

In the large $N$ limit analytic calculations are made possible by the fact that
the functional integral is dominated by the saddle point configuration of the
fields, which in turn can be found by solving the appropriate (saddle point)
equations for the eigenvalues $\phi_i$ of a matrix variable. In practice one
must replace the summations over the index ``$i$'' with an integration in the
variable $\phi$, weighted by an eigenvalue density $\rho (\phi) =
\frac{1}{N}\frac{\rmd i}{\rmd\phi}$, normalized to $\int \rmd\phi\, \rho (\phi) =
1$.

Explicit eigenvalue densities have been found for the infinite volume case
(equivalent to the single plaquette) \cite{Gross:1980he}, and for the chiral
chains with $P = 2,3,4$ \cite{Brower:1980rp, Brower:1980vm}, and the
corresponding free energies have been computed. In all cases a third order
phase transition is present, and therefore one needs to know the separate
expressions for the strong and weak coupling eigenvalue distributions.  As we
saw in the previous sections, as far as we are interested in the topological
susceptibility (or in other properties related to the behaviour of the free
energy close to $\theta=0$) we can use the single plaquette model to compute
values in the thermodynamic limit.

In the single plaquette model the transition occurs at $\beta_c = \frac{1}{2}$
and the eigenvalue density is \cite{Gross:1980he}
\begin{equation}\label{eq:rho_gw}
\rho(\phi,\beta)=\left\{\begin{array}{ll} \rho_s(\phi,\beta) & \mathrm{if}\ \beta\le\beta_c\ , |\phi|\le\pi \\
\rho_w(\phi,\beta) & \mathrm{if}\ \beta>\beta_c\ , |\phi|\le \phi_c \end{array}\right.
\end{equation}
where $\phi_c= 2 \arcsin \sqrt{1/(2\beta)}$ and
\begin{align}
& \rho_s(\phi, \beta) = \frac{1}{2 \pi} \Bigl( 1 + 2 \beta \cos \phi \Bigr) \\
& \rho_w (\phi, \beta) = \frac{2 \beta}{\pi } \cos \frac{\phi}{2}\Bigl(  \frac{1}{2 \beta}- \sin^2 \frac{\phi}{2}  \Bigr)^{\frac{1}{2}} \ .
\end{align}
In order to extend these results to the evaluation of the large $N$ limit of
the topological susceptibility at infinite volume we must replace the saddle
point equation introduced in \cite{Gross:1980he} with
\begin{equation} 
\mathrm{P} \int_{-\phi_c}^{+\phi_c} \rho (\phi',\beta) \cot {\phi - \phi' \over 2} \rmd\phi' - 2 \beta \sin \phi + i {\hat{\theta} \over 2 \pi} = 0\ ,
\end{equation}
where we introduced the scaling variable $\hat{\theta} = \theta/N$ in order to
obtain a consistent large $N$ limit, in analogy with the procedure adopted in
\cite{Bonati:2016tvi, Rossi:2016uce} following the original proposal by
Witten \cite{Witten:1980sp}. We may introduce in the saddle point equation the
{\it Ansatz}
\begin{equation}
\rho(\phi,\beta) = \rho_0(\phi,\beta) + i {\hat{\theta} \over 2 \pi} \rho_1(\phi,\beta),
\end{equation}
where $\rho_0(\phi,\beta)$ is the  eigenvalue density Eq.~\eqref{eq:rho_gw} found in
\cite{Gross:1980he}, while $\rho_1(\phi,\beta)$ must be an odd
function of $\phi$ satisfying the equation 
\begin{equation}\label{eq:sp_phi1}
\mathrm{P} \int_{-\phi_c}^{+\phi_c} \rho_1 (\phi',\beta) \cot {\phi - \phi' \over 2} \rmd\phi' + 1 = 0.
\end{equation}
If we denote by $\mathcal{F}(\beta,\theta)$ the free energy of the system, its
$\theta$-dependent part $F(\beta,\theta)\equiv
\mathcal{F}(\beta,\theta)-\mathcal{F}(\beta,0)$ is therefore\footnote{This
expression is clearly non $2\pi$-periodic in $\theta$, as a consequence of
the use of the single plaquette model.}
\begin{equation}
F(\beta,\theta)= -\frac{1}{2}\Bigl({\theta \over 2 \pi} \Bigr)^2 \int_{-\phi_c}^{\phi_c} \rho_1 (\phi,\beta)\, \phi\, \rmd\phi\	,
\end{equation}
with the factor $1/2$ coming from the partial cancellation of the two terms in
the free energy that are quadratic in $\theta$, i.e. the $\theta$-term and the
term coming from the Haar measure.  In the large $N$ limit the above expression
is finite while all contributions of higher order in $\theta$ are depressed by
powers of $1/N$. Hence we immediately obtain the large $N$ relationship
\begin{equation}\label{eq:chit_rho1}
\chi_t (N, \beta, 1) \rightarrow {1 \over 4 \pi^2}  \int_{-\phi_c}^{\phi_c} \rho_1 (\phi,\beta) \,\phi \,\rmd\phi.
\end{equation}

\begin{figure}[t] 
\centering 
\includegraphics[width=0.9\columnwidth, clip]{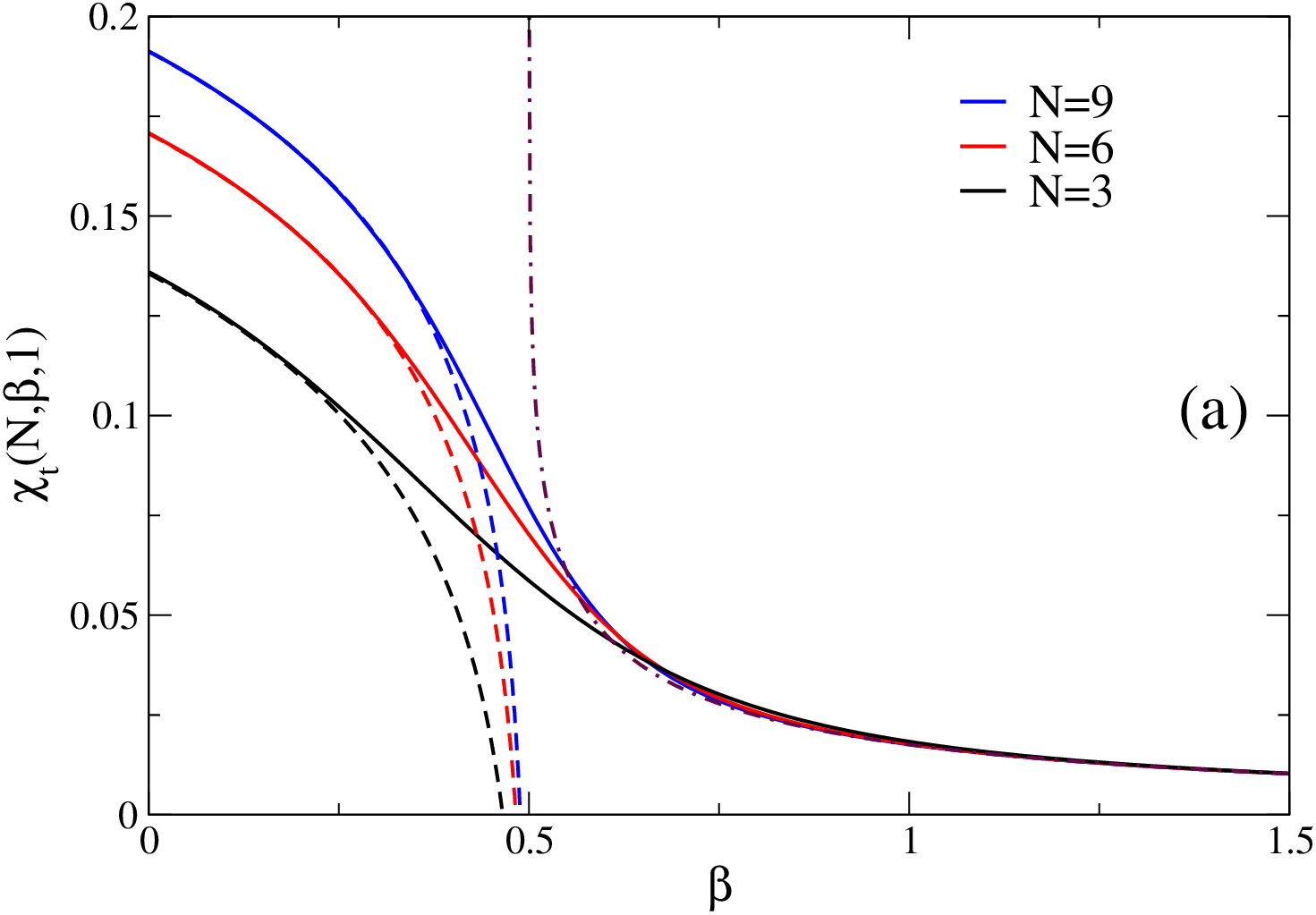}
\includegraphics[width=0.9\columnwidth, clip]{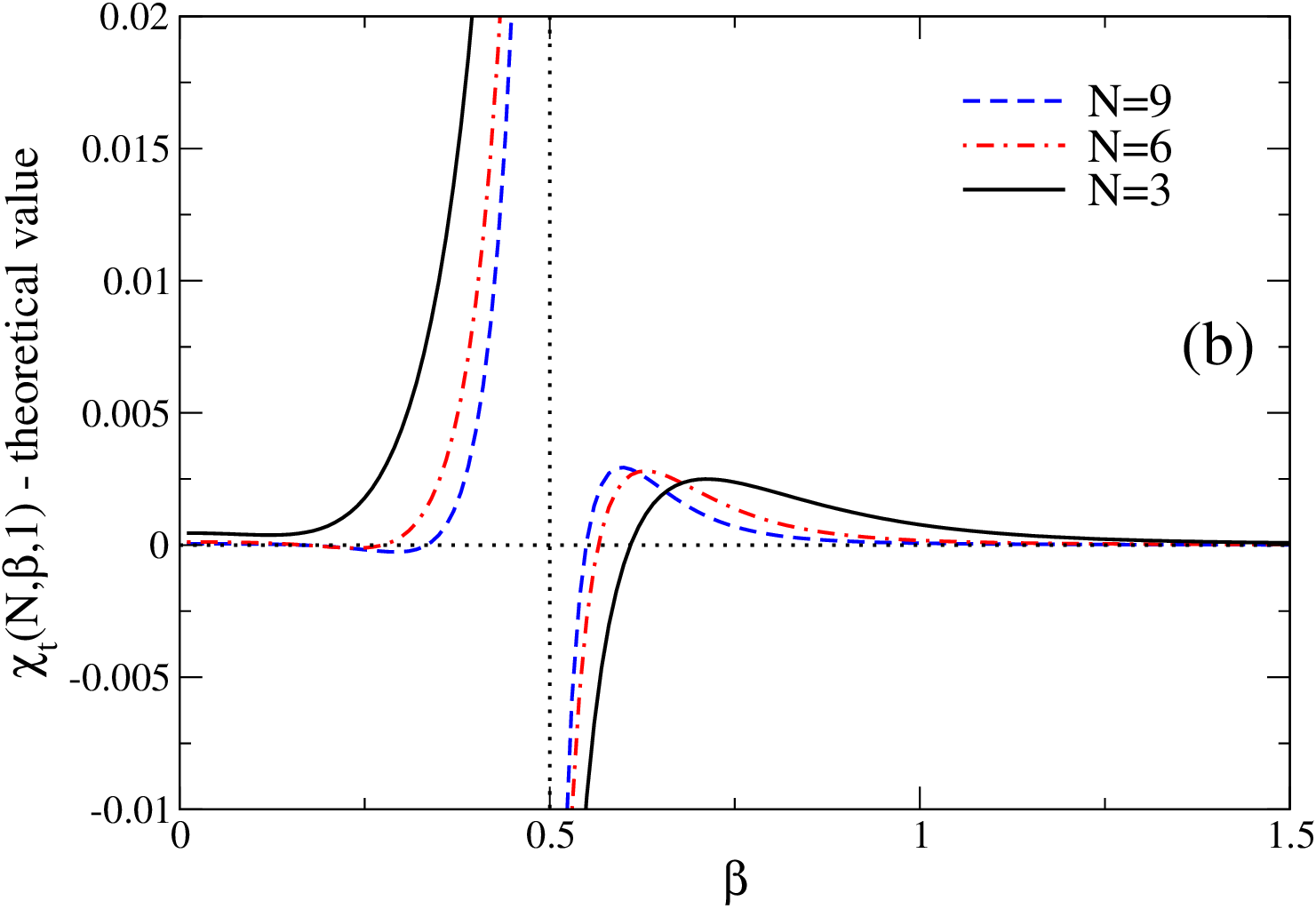}
\caption{(a) Comparison of numerical data obtained by using
Eq.~\eqref{eq:chiNinfiniteV} (solid lines) and the leading order large $N$
theoretical predictions, which is given by Eq.~\eqref{eq:chis_1},
\eqref{eq:chis_2} and \eqref{eq:chis_3} for $\beta<1/2$ (dashed lines) and by
Eq.~\eqref{eq:chiw} for $\beta>1/2$ (dotted-dashed line). (b) Deviations of
numerical data from their expected asymptotic behaviour.} 
\label{fig:largeN}
\end{figure}

Notice that the equation defining $\rho_1(\phi,\beta)$ may depend on $\beta$ only
through the limits of the integration domain, which in turn should not change with
respect to the domain of $\rho_0$, because all change in $\phi_c$ would be
depressed by a power of $1/N$. This observation implies that special care will
be needed in the strong coupling region, because  $\phi_c = \pi$ with no
apparent dependence on $\beta$, but the formal solution for $\rho_{1,s}$ is
\begin{equation}
\rho_{1,s}(\phi,\beta) = \frac{1}{2\pi} \tan {\phi \over 2}\ , \quad \beta <{1 \over 2}\ ,
\end{equation}
implying a nonintegrable singularity around $\pm \pi$. It is easy to get
convinced  that the resulting singular behavior may be parametrized by
\begin{equation}\label{eq:chis_1}
\chi_{t,s} (N,\beta, 1) \rightarrow \chi_{t,s}(N, 0, 1) + f(\beta)\ ,
\end{equation}
where 
\begin{equation}\label{eq:chis_2}
\begin{aligned}
\chi_{t,s} (N, 0, 1) & = {N \over 12} - {1 \over 2 \pi^2} \sum_{k=1}^{N} {N-k \over k^2} \rightarrow\\
&\rightarrow {1 \over 2 \pi^2} (\ln N + \gamma_E + 1) + O(N^{-1})
\end{aligned}
\end{equation}
and $f(\beta)$ is a regular function connected to the $\beta$ dependent cutoff
scale, which is in turn related to the behavior of the density $\rho_{0,s}$ in the
proximity of $\pm \pi$.  On these grounds, since $\rho_{0,s} \sim (1- 2\beta)$ when
$\phi \rightarrow \pm \pi$, we find 
\begin{equation}\label{eq:chis_3}
f(\beta) = {1 \over 2 \pi^2} \ln (1- 2\beta),
\end{equation}
which shows the correct $\beta \rightarrow 0$ limit and exhibits a divergence
in the limit $\beta \rightarrow 1/2$, as required in order to match the weak
coupling behavior.

In the weak coupling regime $\beta>1/2$ the solution of Eq.~\eqref{eq:sp_phi1} is
\begin{equation}
\rho_{1,w}(\phi,\beta)=\frac{1}{2\pi}\frac{\sin(\phi/2)}{\sqrt{\frac{1}{2\beta}-\sin^2(\phi/2)}}\ ,
\end{equation}
the integral in Eq.~\eqref{eq:chit_rho1} is convergent and we get
(using Eq.~3.842.2 of \cite{GradshteynRyzhik})
\begin{equation}\label{eq:chiw}
\chi_{t,w} (N,\beta, 1)  \rightarrow 
-{1 \over 4 \pi^2} \ln \left(1 - {1 \over 2 \beta}\right)\ .
\end{equation}
This result can be easily obtained also without explicitly solving the saddle
point equation, because from the definition of the topological charge we have
\begin{equation}
F(\beta, 2 \pi\ell) = \ln \langle \det U_p^{\ell}\rangle 
\end{equation}
and in \cite{Rossi:1982vw} it has been proven that, at $N=\infty$ in the weak
coupling phase of the single plaquette model, we have
\begin{equation}\label{eq:det}
\langle \det U_p\rangle =  \left(1 - {1 \over 2 \beta}\right)^{1 \over 2}\ ;
\end{equation}
this was further strengthened in \cite{Aneva:1983za} by showing that 
\begin{equation}\label{eq:detl}
\langle \det U_p^{\ell}\rangle =  \left(1 - {1 \over 2 \beta}\right)^{\ell^2/2}\ .
\end{equation}
Hence we may establish the relationship, holding for all $\ell$ and $\beta > 1/2$
\begin{equation}
F(\beta, 2 \pi\ell ) = -\frac{\ell^2}{2}\int_{-\phi_c}^{\phi_c} \rho_1 (\phi,\beta)\, \phi \,d\phi = {\ell^2 \over 2} \ln \left(1 - {1 \over 2 \beta}\right),
\end{equation}
implying immediately
\begin{equation}
\chi_{t,w} (N,\beta, 1)  \rightarrow -{1 \over 4 \pi^2} \ln \left(1 - {1 \over 2 \beta}\right)\ .
\end{equation}
This result reproduces the correct large $\beta$ behavior of the susceptibility
and shows a divergence for $\beta \rightarrow 1/2$, needed in order to match
the strong coupling behavior.  Notice that, due to the singularity in $N$, this
argument could not be applied to the strong coupling phase, where it is known
that $\ln \langle\det U_p\rangle$ is proportional to $N$ and behaves like $\ln \beta$ when
$\beta \rightarrow 0$ \cite{Green:1980bs, Green:1981mx,Rossi:1982vw}.

The numerical evaluation of $\chi_t$, even for quite small values of $N$, shows
surprisingly good agreement with the above predictions, as shown in
Fig.~\eqref{fig:largeN}.

\begin{figure}[t] 
\centering 
\includegraphics[width=0.9\columnwidth, clip]{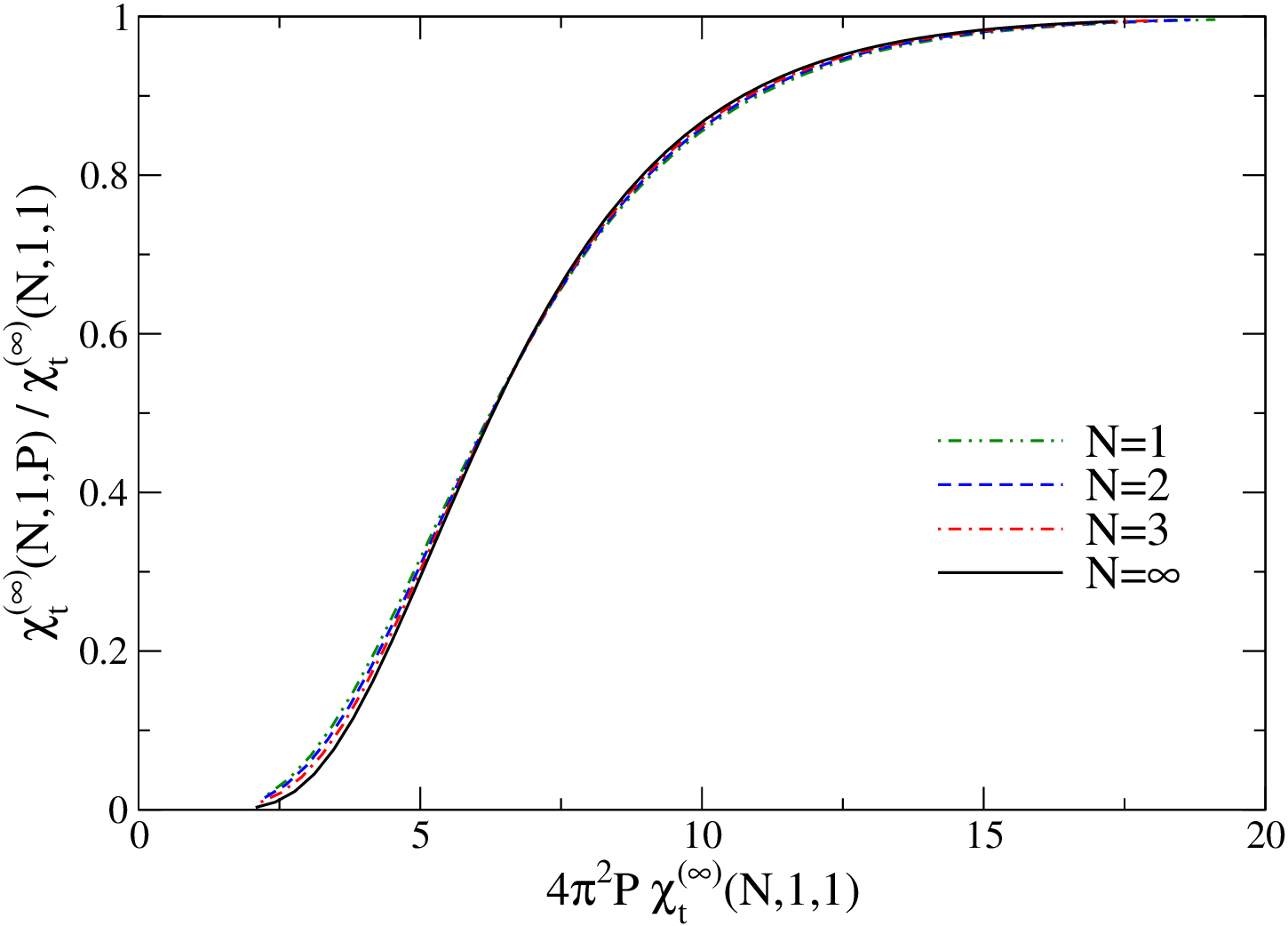}
\caption{Large $N$ scaling of the topological susceptibility at $g=\infty$ for
several values of $P$ at coupling $\beta=1$.}  
\label{fig:finitev_ginfty}
\end{figure}

The above results are restricted to the infinite volume version of the models,
but they may be employed in the $g \rightarrow \infty$ limit in order to obtain
for this case expressions holding also in the finite volume large $N$ limit, at
least in the weak coupling regime. Indeed by trivially  extending
Eq.~\eqref{eq:detl} to include the dependence on $\theta$ and substituting the
results in Eqs.~\eqref{eq:largeg}-\eqref{eq:largeg2} one easily obtains for
large $N$:
\begin{equation}
Z_{\theta}^{(\infty)}(N,\beta,P) \rightarrow [\tilde{\beta}_0(N,\beta)]^P 
\sum_l \left(1 -\frac{1}{2 \beta}\right)^{{P \over 2}\left(l+\frac{\theta}{2\pi}\right)^2}\ .
\end{equation}
This expression can be rewritten in the form
\begin{equation}
Z_{\theta}^{(\infty)}(N,\beta,P) \rightarrow [\tilde{\beta}_0(N,\beta)]^P Z_{\theta}(1,X)\ , 
\end{equation}
where $Z_{\theta}(1,X)$ is the $U(1)$ partition function of the single
plaquette model (see Sec.~\ref{sec:N1}) and therefore
\begin{equation}
\frac{\chi_{t,w}(\infty, \beta, P)}{\chi_{t,w}(\infty, \beta, 1)}=1+2X\frac{\partial}{\partial X} Z_0(1,X)\ ,
\end{equation}
where now
\begin{equation}
X=4\pi^2 P \chi_{t,w}(\infty,\beta,1)\ .
\end{equation}
Here $\chi_{t,w}(\infty,\beta,1)$ is the value Eq.~\eqref{eq:chiw} of the large
$N$ limit of the topological susceptibility in the weak coupling regime, from
which we may appreciate that in the continuum limit $X\to\frac{P}{2\beta}$.  It
is worth noticing that very precocious large $N$ scaling is obtained when
studying $\chi_{t,w}^{(\infty)}(N, \beta,P)/\chi_{t,w}^{(\infty)}(N,\beta,1)$ as a
function of the dimensionless variable $4\pi^2 P \chi_{t,w}(N,\beta,1)$, which
is the finite-$N$ analogous of $X$, see Fig.~\ref{fig:finitev_ginfty}.
This is analogous to what was previously observed in the case of $U(1)$, shown in
Fig.~\ref{fig:u1_finiteV}.

Another important comment concerns the dependence of the large $N$ finite
volume susceptibility on $g$. It is possible to show that the same results
holds true not only for $g\to\infty$, but also for all $g>1$ values, because
representations with $d_r>1$ get suppressed as $N\to\infty$ (see
Eq.~\eqref{eq:weights}). On the other hand it can not hold in the case $g=0$,
since we know that $\chi_t^{(0)}(N,\beta,2)=0$ for all $N$ and, as a
consequence, it vanishes also in the $N\to\infty$ limit. 

By generalizing to general $g$ the arguments put forward by Douglas and Kazakov
\cite{Douglas:1993iia} one may argue that the finite area transition they found
is present only in the $g=0$ case, and it would be interesting to investigate
whether this transition may affect the topological susceptibility.

\section{Conclusions}

In this paper we studied the $\theta$ dependence of two dimensional gauge
theories, providing explicit expressions for the topological susceptibility in
the most general setting, i.e. at finite volume, finite lattice spacing
and for a generic topology of the space-time manifold. 

These expressions can be simplified in several different ways by restricting to
more specific cases. In particular we analyzed the thermodynamic limit at fixed
('t Hooft) coupling and the continuum limit at fixed dimensionless volume, the
case of the abelian $U(1)$ theory being particularly simple. We finally
addressed the large $N$ limit of the results obtained at infinite volume,
showing that the large $N$ behaviour of the topological susceptibility is
completely different for $\beta<1/2$ and for $\beta>1/2$. These two regions
correspond to the strong and weak coupling phases of the $N=\infty$ theory,
separated by the Gross-Witten-Wadia transition.

From the practical point of view our results can be useful to benchmark, in two
dimensional gauge theories, new Monte-Carlo algorithms specifically targeted at
improving the decorrelation of topological modes in lattice gauge theories.
From the theoretical side the most significant results obtained are probably the
determination of the continuum $\theta-$dependent partition function on a
manifold of arbitrary genus and the large $N$ limit (at infinite volume) of the
topological susceptibility for arbitrary coupling.

A remarkable aspect of our large $N$ computation is the fact that the $\theta$
term is sub-leading in the action, but nevertheless we have been able to
compute the topological susceptibility at large $N$ using the saddle-point
approximation method, whose range of applicability is typically restricted to
leading order computations. This is analogous to what has been done in
\cite{Rossi:2016uce} for two dimensional $CP^{N-1}$ models, but the present
case is probably more surprising since the topological susceptibility does not
vanish in the large $N$ limit. 

Putting together the two arguments presented in Sec.~\ref{sec:largeN} to
justify Eq.~\eqref{eq:chiw} we obtain a new and completely independent proof of
Eq.~\eqref{eq:det}, suggested in \cite{Green:1980bs, Green:1981mx} and proven
in \cite{Rossi:1982vw}, and of Eq.~\eqref{eq:detl}, proven in
\cite{Aneva:1983za}. A natural question is whether the new proof can be
extended to other cases that were not tractable with the previously known
methods.

\emph{Acknowledgement} It is a pleasure to thank M.~D'Elia for useful discussions.

\end{document}